\documentclass[preprint,epsf,aps,floats]{revtex4}
\usepackage{amsmath,amssymb,bm,epsfig,psfrag}
\renewcommand\d{\partial}

\newcommand{\be}{\begin{equation}}
\newcommand{\ee}{\end{equation}}
\newcommand{\ber}{\begin{eqnarray}}
\newcommand{\eer}{\end{eqnarray}}

\def\Dsl{\,\raise.15ex \hbox{/}\mkern-12.8mu D}

\begin{document}

\title{Effective Lagrangian of unitary Fermi gas from $\varepsilon$ expansion}
\author{Andrei~Kryjevski}
\affiliation{Nuclear Theory Center, Indiana University, Bloomington, IN, 47408 and
Washington University in St. Louis, Department of Physics, St. Louis, MO 63130}

\begin{abstract}
Using $\varepsilon$ expansion technique proposed in \cite{Nishida:2006br} we derive an effective
Lagrangian (Ginzburg-Landau-like functional) of the degenerate unitary Fermi gas to the next-to-leading (NLO) order in
$\varepsilon.$
It is demonstrated that for many realistic situations it is sufficient to retain
leading order (LO) terms in the derivative expansion.
The functional is used to study
vortex structure in the symmetric gas,
and interface between normal and superfluid phases in the polarized gas.
The resulting surface free energy is about four times larger than the value previously quoted in the literature.
\end{abstract}

\date{\today}

\maketitle

\section{Introduction}
There has been considerable effort, both theoretical and experimental, to understand properties
of dilute strongly interacting cold fermion systems realized, for example, in the experiments on
trapped cold atomic gases, and in dilute neutron matter encountered in
the neutron star crust \cite{ohara-2002-2002,regal-2004,bartenstein-2004-92,zwierlein-2005-435,zwierlein-2006-311,partridge-2005,partridge-2006-97}. The case of infinite
scattering length and zero effective range of the two body interactions has been named "unitary Fermi
gas" and is particularly interesting due to its universal properties.

Inspired by work of Nussinov and Nussinov \cite{Nussinov:2004nn}, an analytical technique similar
to the $\varepsilon$ expansion in the theory of critical phenomena has recently been proposed
\cite{Nishida:2006br}. 
Comparison with the results of the state-of-the-art Monte Carlo simulations and with experimental results
suggested that already at the next-to-leading order (NLO) the $\varepsilon$ expansion might be a useful
tool in the description of this system \cite{Nishida:2006br,Nishida:2006rp,Rupak:2006jj}.
While subsequent investigation raised doubts about the convergence of the series
\cite{Arnold:2006fr},
one obvious way to check usefulness of the technique is to make predictions for various observable quantities at NLO.
If the trend suggested by the initial NLO results of \cite{Nishida:2006br,Nishida:2006rp,Rupak:2006jj}
were to hold, it would serve as a strong encouragement for further investigation. 

So, as a step in this direction, in this article we derive an effective Lagrangian of the degenerate
unitary Fermi gas in derivative expansion, and to the next-to-leading order (NLO) in $\varepsilon$ which is the subject of Section 2. The
functional is then applied to two spatially inhomogeneous phenomena in the unitary Fermi gas:
single vortex structure and interface between normal and
superfluid phases of the imbalanced gas at the critical polarization (Sections 3 and 4). In this paper predictions for $d=3$
will be made by just setting $\varepsilon=1$ in the end of NLO calculations. More sophisticated extrapolations
to $d=3$ \cite{Nishida:2006rp,Arnold:2006fr} are left to future work. Some further applications the effective
Lagrangian functional are discussed in Section 5.

\section{Effective Lagrangian to NLO in $\varepsilon$}

The dynamics of the system is governed by the Hamiltonian
\ber
{\rm H}=
\sum_{\alpha=\uparrow,\downarrow}\int_{\vec{x}}\psi^{\dagger}_{\alpha}\left(-\frac{\nabla^2}{2m}-\mu_{\alpha}\right)\psi_{\alpha}-
c_0\,\psi^{\dagger}_{\uparrow}\psi_{\uparrow}\psi^{\dagger}_{\downarrow}\psi_{\downarrow},
\label{HH}
\eer
the corresponding Lagrangian density is given by
\ber
\mathcal{L} &=&
\psi^{\dagger}\left(i\d_t + \frac{\nabla^2}{2m}+ \mu\right)\psi \nonumber \\
 &+& c_0\,\psi^{\dagger}_{\uparrow}\psi_{\uparrow}\psi^{\dagger}_{\downarrow}\psi_{\downarrow}
\label{LL}
\eer
where
$\psi=(\psi_\uparrow,\psi_\downarrow)^T,$ is spin-1/2 fermion field,
$\mu={\rm diag}(\mu_\uparrow,\mu_\downarrow)$ are the chemical potentials for the two pairing species.

The coupling $c_0$ is to be tuned to reproduce a desirable two body scattering length,
$a.$ In this article we will concentrate on the unitary regime, $a=\infty,$ which in the dimensional
regularization corresponds to $1/c_0=0$ \cite{Nishida:2006br}.
It is possible to use $\varepsilon$ expansion to describe system near unitarity \cite{Nishida:2006eu},
and the results of this paper may be extended to the near-unitary regime. This is left for future work.

A convenient tool that we will use to study inhomogeneous configurations of the unitary Fermi gas is the
effective Lagrangian of the theory (\ref{LL}) (see, for example, Ch. 11 of \cite{Peskin:1995ps} for an introduction). 
The zero temperature effective action functional
$\Gamma[\Phi(x),\mu,\delta \mu]$ of
the order parameter $\Phi(x)\propto\langle\psi(x)_{\uparrow}\,\psi(x)_{\downarrow}\rangle,$
where $\langle...\rangle$ means expectation value in the ground state, is defined by the following
\ber
\Gamma[\Phi(x),\mu,\delta\mu]=-\Omega[J(x),\mu,\delta\mu]-\left(\int_{x}{J^*}(x)\,\Phi(x)+h.c.\right),~~x\equiv(t,\vec{x}),
\label{Seff}
\eer
with the source $J(x)$ satisfying $$\frac{\delta}{\delta J^*(x)}\Omega[J(x),\mu,\delta\mu]=-\Phi(x)$$
and
\ber
e^{-i\,\Omega[J]}
=\int {\rm D \phi}\,{\rm D \phi^*}\,
{\rm det}{\left(\begin{array}{ccc}
i\,\partial_t+\frac{\nabla^2}{2\,m}+\mu_{\uparrow}& {\phi^*}(x) \\ {\phi}(x)&
i\,\partial_t-\frac{\nabla^2}{2\,m}-\mu_{\downarrow}\\\end{array}\right)}
{\rm exp}~\,i\left(\int{J^*}\phi+J\,\phi^*\right),
\label{Omega}
\eer
where $\phi(x)$ is the auxiliary di-fermion field introduced by the Hubbard-Stratonovich
transformation \cite{Nishida:2006br, Nishida:2006eu}; $1/c_0=0$ has been used;
we have introduced $\mu=(\mu_{\uparrow}+\mu_{\downarrow})/2,~\delta\mu=(\mu_{\uparrow}-\mu_{\downarrow})/2.$
The theory is defined in $d=4-\varepsilon$ spatial dimensions and the calculations are performed using perturbation theory in $\varepsilon$ about $d=4$ \cite{Nishida:2006br, Nishida:2006eu}.

The objective is to integrate out $\phi$ in (\ref{Omega}) to a given order in $\varepsilon$ in the presence of
$J(x)$ and then perform the Legendre transformation.
To approximate
$\Gamma[\Phi(x),\mu,\delta\mu]$ which is, in general, non-local,
we will attempt an expansion in derivatives of $\Phi(x),$ and, so, will assume
\ber
\Gamma[\Phi(x),\mu,\delta\mu]=\int_x {\rm L}_{eff}[\Phi(x),\mu,\delta\mu]\equiv\int_x -{\rm V}_{eff}(\Phi(x),\mu,\delta\mu)+ {\rm derivative}~{\rm terms}.
\label{GammaLeff}
\eer
So, in the case of a homogeneous medium effective Lagrangian reduces to effective potential,
${\rm L}_{eff}[\Phi,\mu,\delta\mu]=-\rm{V}_{eff}(\Phi,\mu)$ for $\Phi={\rm const}.$ The effective potential for symmetric (near)
unitary Fermi gas ($\delta\mu=0$) has been calculated to NLO in $\varepsilon$ in \cite{Nishida:2006br,Nishida:2006eu}; the calculation has been extended to NNLO in $\varepsilon$ by Arnold, Drut and Son \cite{Arnold:2006fr}. The NLO effective potential for the polarized case ($\delta\mu \neq 0$) was derived in \cite{Rupak:2006et,Nishida:2006eu}. As already stated, in this article we will be working to NLO in $\varepsilon$ at unitarity. Within derivative expansion approach the task is to augment the potential with derivative terms.
In the following we are going to argue that at least the leading order (LO) derivative terms may be determined by
imposing gauge invariance on the theory, which allows one to avoid performing explicit Legendre transformation.
The derivative expansion approach is reminiscent of the Ginzburg-Landau functional construction,
but here we do not assume being in the vicinity of a second order phase transition. Also
here we have an (approximate) control over the microscopic theory and are able to calculate the coefficients
in the functional.
Note that if one were to fail to find a sensible derivative expansion where only few terms have to be kept to achieve given accuracy, then the effective action would indeed be non-local and a more general technique (such as the standard Bogoliubov-de Gennes method \cite{Degennes:1966fr}) would have to be employed. But in the following we are going to argue that for a few realistic situations the derivative expansion is related to the $\varepsilon$ expansion and, so, a finite number of derivative terms has to be kept to a given order in $\varepsilon.$

Once the functional (\ref{GammaLeff}) is known, the ground state is characterized by the equation of motion
\ber
\frac{\delta}{\delta \Phi(x)}{\rm L}_{eff}[\Phi(x),\mu,\delta\mu]=0.
\label{eomPhi}
\eer
Then, for example, free energy and particle numbers are given by
\ber
&&{\rm F}(\mu,\delta\mu)=-\int{\rm d}^dx\,{\rm L}_{eff}[\hat{\Phi}(x),\mu,\delta\mu]\nonumber \\
&&{\rm N_1+N_2}=-\frac{\partial {\rm F}(\mu,\delta\mu)}{\partial \mu}\qquad{\rm N_1-N_2}=
-\frac{\partial {\rm F}(\mu,\delta\mu)}{\partial \delta\mu},
\label{N1N2}
\eer
where $\hat{\Phi}(x)$ is a solution of (\ref{eomPhi}) and ${\rm N_1,N_2}$ are the particle numbers of the two species.

\subsection{Effective Lagrangian for symmetric unitary Fermi gas ($\delta\mu=0$) to NLO in $\varepsilon$}

Let us begin by considering the case of unpolarized unitary Fermi gas ($\delta\mu=0$). The effective potential (that is, non-derivative terms of the effective action) for unitary Fermi gas has been calculated to ${\cal{O}}(\varepsilon)$ by Nishida and Son
~\cite{Nishida:2006br}
\ber
 {\rm{V}}_{eff}(\Phi(x),\mu)&=&
 \left(\frac{m\,|\Phi(x)|}{2\pi}\right)^{d/2}\,\frac{|\Phi(x)|}3\left[1+\frac{7-3(\gamma+\ln2)}6\,\varepsilon-3\,C\varepsilon\right]-
\nonumber \\
 &-&\left(\frac{m\,|\Phi(x)|}{2\pi}\right)^{d/2}\,\frac\mu\varepsilon
\left[1+\frac{1-2(\gamma-\ln2)}4\,\varepsilon\right],
\label{Veff}
\eer
where $\gamma\approx0.57722$, is the Euler-Mascheroni constant and $C\approx0.14424.$ Minimization yields
\ber
\phi_0=2\,\mu_0\left(1+\varepsilon(3\,C-1+{\rm Log}\,2)\right),~~\mu_0=\frac{\mu}{\varepsilon}\sim 1,
\label{phi0}
\eer
the NLO homogeneous medium superfluid order parameter.
\subsubsection{The Leading Order Derivative Terms}
General coordinate and conformal invariance has been used by Son and Wingate to constrain low energy
effective Lagrangian of the unitary Fermi gas for the phonon field (the phase of di-fermion condensate) 
\cite{Son:2005rv}. However, in the situations that will be dealt with in this paper it will be more convenient to 
have an effective Lagrangian depending on both phase and magnitude of the order parameter. 

We will argue that within the framework of $\varepsilon$ expansion just gauge invariance will be sufficient 
to determine at least the leading order (LO) derivative terms. The basic idea is to consider a generalization 
of the theory (\ref{LL}) by gauging the $U(1)$ particle symmetry and imposing gauge invariance. If a certain 
relationship is true for the effective Lagrangian of the generalized theory, then it also has to hold for the original theory.

Our theory (\ref{LL}) is invariant under $U(1)_{\uparrow}\times U(1)_{\downarrow}\equiv U(1)_{V}\times U(1)_{A}$ which are particle number 
symmetries for the spin up and down species. To obtain LO derivative terms for both polarized and unpolarized cases it is sufficient to 
gauge $U(1)_{V}$ and require that the Lagrangian be invariant in the presence of an arbitrary external gauge field $A_{\mu}=(A_0,\vec A).$
Then Lagrangian (\ref{LL}) becomes
\ber
{\rm L}(A_0(x),\vec{A}(x))&=&{\psi}^{\dagger}\,(i\,\d_t-A_0)\,\psi-\frac{1}{2\,m}\,(\vec{\nabla}\psi^{\dagger}-i\,\vec{A}\,\psi^{\dagger})\cdot(\vec{\nabla}\psi+i\,\vec{A}\,\psi)\nonumber \\
 &+& c_0\,\psi^{\dagger}_{\uparrow}\psi_{\uparrow}\psi^{\dagger}_{\downarrow}\psi_{\downarrow}.
\label{gaugedL}
\eer
Under an infinitesimal $U(1)$ transformation, ${\rm exp}~i\alpha(x),$ the fields transform as
\ber
&&\psi\rightarrow (1+i\alpha(x))\psi,\,\phi\rightarrow (1+2i\alpha(x))\phi \nonumber \\
&&A_0 \rightarrow A_0 - \partial_t \alpha(x),\,\vec{A} \rightarrow \vec{A} - \vec{\nabla} \alpha(x).
\label{transformation}
\eer
For example, 
the case of a gas trapped in a potential ${\rm U}(\vec{x})$ is represented by
$A_0=-\mu+{\rm U}(\vec{x})$ and $\vec{A}=0.$
To determine the LO derivative terms it will be sufficient to consider effective potential 
in the presence of $A_{\mu}=const.$

To determine the leading order (LO) derivative terms we observe that
\begin{enumerate}
\item
the ${\vec A}^2$ term appears in the bare Lagrangian in the sum with $A_0$ as
\ber
-{\psi}^{\dagger}\,\left(A_0+\frac{{\vec A}^2}{2\,m}\right)\,\psi;
\label{A0Acouple_to_density}
\eer

\item
the
$\vec{A}\cdot\,i\,\psi^{\dagger}\vec{\nabla}\,\psi/2\,m+h.c.$ term does not contribute to the
coefficient of the ${\vec A}^2$ term in the effective action, ${\it {i.e.}}$ to the Meissner mass of the $U(1)$ gauge boson;

\item
by gauge and rotational invariance the lowest order effective action derivative term has the form
\ber
Z_1(\Phi)\,\Phi^{*}(i\,\partial_t-2\,A_0)\,\Phi&-&
Z_2(\Phi)(\vec{\nabla}\Phi^{*}-2\,i\,\vec{A}\,\Phi^{*})\cdot(\vec{\nabla}\Phi+2\,i\,\vec{A}\,\Phi)=\nonumber\\&=&
Z_1(\Phi)\,\Phi^{*}(i\,\partial_t-2\,A_0)\,\Phi-
Z_2(\Phi)(|\vec{\nabla}\Phi|^2+4\,{\vec A}^2\,|\Phi|^2)+...\,,
\label{DphiDphi}
\eer
where $Z_1(\Phi)$ and $Z_2(\Phi)$ are some functions.
\end{enumerate}

Functions $Z_1(\Phi)$ and $Z_2(\Phi)$ are to be determined from the coefficient at
$A_0+\frac{{\vec A}^2}{2\,m}$ in the effective potential in
the background of constant gauge field $A_{\mu}=(A_0,\vec A).$ But the calculation for $A_0=-\mu$
has already been performed by Nishida and Son with the result shown in (\ref{Veff}).
Then the LO derivative terms are
\ber
 \left(\Phi^{*}\,i\,\partial_t\,\Phi-\frac{1}{4\,m}|\vec{\nabla}\Phi|^2\right)
\frac{1}{2\,|\Phi|^2\,\varepsilon}
\left(1+\frac{1-2(\gamma-\ln2)}4\,\varepsilon\right)\left(\frac{m\,|\Phi|}{2\pi}\right)^{d/2}.
\label{derivative_terms}
\eer

Note that the coefficient at $({\vec A}^2)^2$ term in ${\rm V}_{eff}(A_0,{\vec A})$ is given by a convergent integral and is ${\cal O}(1).$ This will be used in the next section.
\subsubsection{The Higher Order Derivative Terms}
Now let us consider the higher derivative terms. 
Gauge and rotational symmetries are not sufficient to specify derivative terms unambiguously, but as will become clear from the
following, it will suffice to determine the order of the derivative coefficients in $\varepsilon.$
For simplicity, we will concentrate on time independent configurations, but a similar argument may be made for the terms with time
derivatives.

\begin{figure}[t]
\centering{
\begin{psfrags}
\psfrag{phis}{$\delta\phi(k)^*$}
\psfrag{phi}{$\delta\phi(k)$}
\epsfig{figure=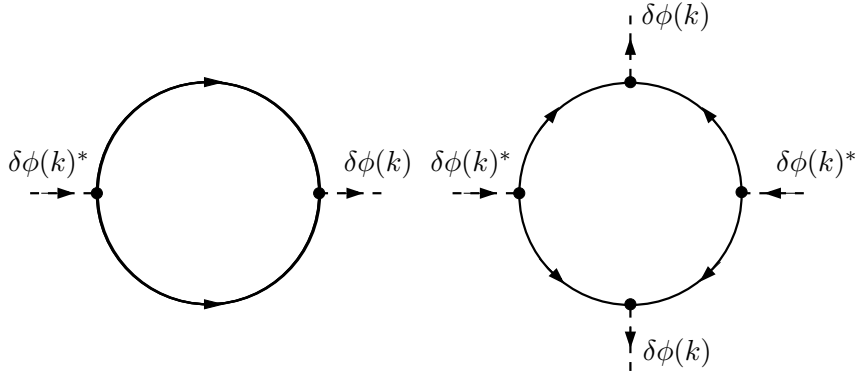, width=.65\textwidth}
\end{psfrags}
}
\caption{Diagrams that produce the LO contributions to the operators in (\ref{Kineffdelta12}). The fermion propagators in the $\phi_0$ background are the solid
lines, the $\delta\phi(\vec{k})$ insertions are the dashed lines. The arrows indicate the direction of particle number flow. Only normal contributions are shown.}
\label{fig:graphs}
\end{figure}

An $n$-derivative term has the form
\ber
\left(\frac{m\,|\Phi|}{2\pi}\right)^{d/2}\,\phi_0\,
Z_{s\,r}\left(\frac{\Phi}{\phi_0}\right)\frac{1}{m^{\frac{n}{2}}\,\phi_0^{\frac{n}{2}}\,L^n}
\frac{|\vec{\nabla}^{r}_{\tilde{x}}\Phi|^s}{|\Phi|^s},
\label{nderivative}
\eer
with various combinations of $r$ and $s \geq 2$ compatible with the rotational invariance and such that $2r\cdot s=n,$ where
$Z_{s\,r}\left(\frac{\Phi}{\phi_0}\right)$
are some functions, $\phi_0$ is the uniform medium order parameter. A dimensionless variable
${\vec{\tilde{x}}}={\vec{x}}/L$ has been introduced with $L$ being the characteristic length scale so that
${\partial}_{\tilde{x}}\Phi(\tilde{x}) \sim 1.$
The terms with two and four spatial derivatives are then
\ber
&-&\phi_0\,\left(\frac{m\,|\Phi|}{2\pi}\right)^{d/2}
\frac{|\vec{\nabla}_{\tilde{x}}\Phi|^2}{8\,m\,\phi_0\,L^2\,|\Phi|^2\,\varepsilon}
\left(1+\frac{1-2(\gamma-\ln2)}4\,\varepsilon\right),
\label{2derivative}
\eer
\ber
\nonumber\\&-&\phi_0\,\left(\frac{m\,|\Phi|}{2\pi}\right)^{d/2}\,\frac{1}{m^2\,\phi_0^2\,L^4}\,
\left(Z_{41}\left(\frac{\Phi}{\phi_0}\right)
\frac{(|\vec{\nabla}_{\tilde{x}}\Phi|^2)^2}{|\Phi|^4}+
Z_{22}\left(\frac{\Phi}{\phi_0}\right)
\frac{|\vec{\nabla}^2_{\tilde{x}}\Phi|^2}{|\Phi|^2}\right)
\label{4derivative}
\eer
where $Z_{41},~Z_{22}$ are unknown functions.
We may determine magnitude of the $Z$'s by considering a weakly inhomogeneous configuration,
$\Phi(\vec x)=\phi_0+\delta\phi(\vec x)$ with $|\delta\phi| \ll \phi_0.$ On one hand, in this case
(\ref{4derivative}) reduces to
\ber
\nonumber\\&-&\phi_0\,\left(\frac{m\,\phi_0}{2\pi}\right)^{d/2}\,\frac{1}{m^2\,\phi_0^2\,L^4}\,
\left(Z_{41}\left(1\right)
\frac{(|\vec{\nabla}_{\tilde{x}}\delta\phi|^2)^2}{{\phi_0}^4}+
Z_{22}\left(1\right)
\frac{|\vec{\nabla}^2_{\tilde{x}}\delta\phi|^2}{{\phi_0}^2}\right)+...
\label{4derivative0}
\eer
where terms of higher order in $\delta\phi$ are omitted.
On the other hand, $\delta\phi$ derivative terms
may be obtained by expanding
(\ref{Omega}) with $J=0$ and $\phi(\vec k)=\phi_0+\delta\phi(\vec k)$ through the 4th order in
$\delta\phi(\vec k)$ and in the transfer momentum, $\vec k.$  This produces the following operators in the
effective Lagrangian
\ber
 \left(\frac{m\,\phi_0}{2\pi}\right)^{d/2}
\left(c_{12}\,\frac{|\vec{\nabla}\delta\phi(x)|^2}{m\,\phi_0^2}+c_{41}\,\frac{(|\vec{\nabla}\delta\phi(x)|^2)^2}{m^2\,\phi_0^5}+c_{22}\,\frac{|\vec{\nabla}^2\delta\phi(x)|^2}{m^2\,\phi_0^3}\right),
\label{Kineffdelta12}
\eer
where $c_{22},\,c_{41}$ and $c_{12}$ are some coefficients.
Shown in Fig.~(\ref{fig:graphs}) are the diagrams that give LO contributions to
(\ref{Kineffdelta12}). The coefficient at $k^2\,|\delta\phi(k)|^2$ is given by a divergent integral which in the dimensional regularization is ${\cal O}(\varepsilon^{-1}),$
while $k^4\,|\delta\phi(k)|^2$ and $k^4\,|\delta\phi(k)|^4$ terms are multiplied by convergent integrals and, thus, have coefficients of
order one. So, we see that $c_{22},\,c_{41}$ are of order one, while $c_{12}$ is ${\cal O}(\varepsilon^{-1}).$

Ignoring an unlikely possibility that a $Z(1)$ and a $Z\left({\Phi}/{\phi_0}\right),~{\Phi}\neq{\phi_0}$ may be of
different order in $\varepsilon$ and
combining this with an observation that the sum of $Z_{41}$ and $Z_{22}$ is given by the
${\cal O}(1)$ coefficient of the $({\vec A}^2)^2$ term of ${\rm{V}_{eff}}(\Phi,A_0,\vec{A}),$ which was already mentioned
in the end of the previous section,
we conclude that $Z_{41}$ and $Z_{22}$ are of order one.

More generally, an operator ${\vec k}^n\,|\delta\phi({\vec k})|^l$ with $n>2$ and $l\geq2$ will have a coefficient given by a convergent integral
${\cal O}(1).$ Then all the higher order derivative terms will have coefficients ${\cal O}(1).$

We observe presence of two length scales in the derivative terms:
\ber
L_{\mu}\sim(\varepsilon\,m\,\phi_0)^{-1/2}\sim (m\,\mu)^{-1/2}\sim\varepsilon^{-1/2},
\label{L1}
\eer
which is proportional to the inter particle separation, and
\ber
L_{sf}\sim\,(m\,\phi_0)^{-1/2}\sim 1,
\label{L2}
\eer
which is superfluidity scale (coherence length).

Note that $L_{\mu}/L_{sf} \sim\varepsilon^{-1/2}\gg 1$ in the $\varepsilon$ expansion.
In the case $L \geq L_{\mu}$ the four and higher order derivative terms are suppressed by at least two powers of
$\varepsilon.$

So, for the situations where one is interested in large enough scale phenomena so that
$L \geq (m\,\mu)^{-1/2}$ and typical time scale $T \geq \mu^{-1}$ the effective Lagrangian of unitary Fermi gas to NLO in the
$\varepsilon$ expansion is
\ber
{\rm{L}}_{eff}[\Phi(x),\mu]&=&
\left[\left(\Phi^{*}\,i\,\partial_t\,\Phi-\frac{1}{4\,m}|\vec{\nabla}\Phi|^2\right)
\frac{1}{2\,|\Phi|^2\,\varepsilon}+\frac\mu\varepsilon\right]
\left(1+\frac{1-2(\gamma-\ln2)}4\,\varepsilon\right)\left(\frac{m\,|\Phi|}{2\pi}\right)^{d/2}-\nonumber \\&-&
\left(\frac{m\,|\Phi|}{2\pi}\right)^{d/2}\,\frac{|\Phi|}3\left[1+\frac{7-3(\gamma+\ln2)}6\,\varepsilon-3\,C\varepsilon\right],
\label{Leff}
\eer
where $\gamma\approx0.57722$, is the Euler-Mascheroni constant and $C\approx0.14424.$

Now let us quote the NLO effective Lagrangian for the polarized unitary Fermi gas $\delta\mu \neq 0$
valid for $L \geq (m\,\mu)^{-1/2}$ and $T \geq \mu^{-1}$ 
(the effective potential was derived in \cite{Rupak:2006et,Nishida:2006eu}).
The derivative terms are deduced by the same line of argument as in the symmetric case.
\ber
{\rm{L}}_{eff}[\Phi,\mu,\delta\mu]&=&\,
\left[\left(\Phi^{*}\,i\,\partial_t\,\Phi-\frac{1}{4\,m}|\vec{\nabla}\Phi|^2\right)
\frac{1}{2\,|\Phi|^2\,\varepsilon}+\frac\mu\varepsilon\right]\left(\frac{m\,|\Phi|}{2\pi}\right)^{d/2}\nonumber\\&\times&
\left(1+\frac{1-2(\gamma-\ln2)}4\,\varepsilon+\frac{\varepsilon}{2}
\Theta(h-1)\left(h\,\sqrt{h^2-1}-\ln(h+\sqrt{h^2-1})\right)\right)-
\nonumber\\&-&
\left(\frac{m\,|\Phi|}{2\pi}\right)^{d/2}\,\frac{|\Phi|}3\left[\left(1+\frac{7-3(\gamma+\ln2)}6\,\varepsilon\right)(1-\Theta(h-1))-3\,C\varepsilon\right]- \nonumber \\
 &-&\left(\frac{m\,|\Phi|}{2\pi}\right)^{d/2}\,\Theta(h-1)
\left[\frac{\varepsilon\,|\Phi|}{24}\,(h-1)^2(h+2)\ln(h^2-1)+
\frac{\delta\mu}{2}\left(1+\frac{\varepsilon}{12}(13-6\,\gamma)\right)\right]-\nonumber \\&-&
\left(\frac{m\,|\Phi|}{2\pi}\right)^{d/2}\,\Theta(h-1)
\left[\frac{\delta\mu^3}{6\,|\Phi|^2}\left(1+\frac{\varepsilon}{12}(11-6\,\gamma)\right)-\frac{\varepsilon\,|\Phi|}{6}\ln(h+1)\right]-\nonumber \\&-&
\left(\frac{m\,|\Phi|}{2\pi}\right)^{d/2}\,\Theta(h-1)\,\varepsilon\,|\Phi|({\cal G}(\Phi)+{\cal H}(\Phi)),~h=\frac{\delta\mu}{|\Phi|}
\label{Leffpolreal}
\eer
where
\ber
{\cal G}(\Phi)&=&\int_0^{\infty}{\rm d}\,x\int_0^{\lambda}{\rm d}\,y\frac{[f(x)-x][f(y)-y]}{f(x)\,f(y)}
[j(x,y)-\sqrt{j(x,y)^2-x\,y}],\nonumber \\
{\cal H}(\Phi)&=&\int_0^{\lambda}{\rm d}\,x\int_{\lambda}^{\infty}{\rm d}\,y\frac{[f(x)+x][f(y)-y]}{f(x)\,f(y)}
[k(x,y)-\sqrt{k(x,y)^2-x\,y}],\nonumber \\
j(x,y)&=&f(x)+f(y)+(x+y)/2,~k(x,y)=f(x)-f(y)-(x+y)/2,~f(x)=\sqrt{x^2+1},\nonumber \\
~\lambda&=&\sqrt{\left({\delta\mu}/{|\Phi|}\right)^2-1}.
\label{GH}
\eer

Let us check if the LO terms of the superfluid mode Lagrangian may be obtained from (\ref{Leff}).
Setting
$\Phi(\vec{x},t)=\rho(\vec{x},t)\,{\rm exp}\,2\,i\,\beta(\vec{x},t)$
and using the solution to the equation of motion for the radial mode $\rho=\rho(\beta;\mu)$ we get
\ber
{\rm{L}}_{eff}[\beta]=
\frac{1}{2}\,\frac{\partial\,n}{\partial\,\mu}\,\left(\partial_t\beta\right)^2-\frac{n\,|\vec{\nabla}\beta(\vec{x})|^2}{2\,m},
\label{betakinterm_x}
\eer
where
\ber
n=\left(\frac{m\,\mu_0}{2\pi\,}\right)^{d/2}\frac{4}{\varepsilon}\left(1+\left[-\frac74+6\,C -\frac{\gamma}2 +
2~{\rm Log}~2\right]\varepsilon\right),\qquad \mu_0=\mu/\varepsilon,
\label{n}
\eer
with $C=0.14424,$
is the NLO number density \cite{Nishida:2006br}. This reproduces previously found LO terms of the superfluid mode Lagrangian \cite{Son:2005rv,son-2006-74}.

Let us comment on the relation of $L_{\mu}$ to the interparticle separation. Using (\ref{n}) we observe that $L_{\mu}\sim L_{int}~\varepsilon^{-1/2-1/d},$ where $L_{int}\sim n^{-1/d}$ is the interparticle separation. Numerically $L_{\mu}\sim L_{int}$ except for 
very small $\varepsilon$ which is not the region we are ultimately interested in, anyway. For example, for $\varepsilon=0.2$ $\varepsilon^{1/2+1/d}\simeq 0.3$ to NLO, while the four derivative term is suppressed by $\varepsilon^2=0.04.$
\section{Unitary Fermi Gas Vortex Structure}

The $U(1)$ particle number symmetry is spontaneously broken in the superfluid ground state of unitary Fermi gas and, therefore, stable vortex configurations are expected and have, in fact, been observed \cite{zwierlein-2005-435}.
So, as a first application of the functional obtained (and as a sensibility check of the formalism) we will investigate the structure of a superfluid vortex in the symmetric 
($\delta\mu=0$) unitary Fermi gas. The problem has been addressed in, ${\it e.g.},$ \cite{bulgac-2003-91} using a different method.

Using Lagrangian (\ref{Leff}) consider a single vortex configuration of unit winding number and set
$\Phi(\vec{x})=\rho(r)\,e^{i\,\theta}$ with $\vec{x}={\{}r,\theta,...{\}}$. We will assume that a vortex filament is two dimensional in
any number of space dimensions. The resulting equation of motion has been solved order by order in $\varepsilon$ to NLO. The LO, NLO
contributions and their sum with $\varepsilon=1,$ the prediction for 3 dimensions, are shown in Fig.~(1).
In the region near the vortex core for $r\sim (m\,\phi_0)^{-1/2}\ll L_{\mu},$ the derivative expansion will break down.
However, since in a vortex $\Phi(r)\rightarrow 0$ as $r\rightarrow 0,$ the error in the contribution to, for example, free energy from the region $r \leq (m\,\phi_0)^{-1/2} $ is insignificant.
The typical size of a single vortex configuration, ${{r}}_0$, defined by
$\rho({{r}}_0)=\rho({{r}}=\infty)/2$ is ${{r}_0}=.43\,({m\,\mu})^{-1/2}$ at LO and
${{r}_0}=.45\,({m\,\mu})^{-1/2}$ at NLO.
\begin{figure}[t]
\centering{
\begin{psfrags}
\psfrag{a}{$4\,\sqrt{m\,\mu}\,{r}$}
\psfrag{b}{${\rho}/{2\,\mu_0}$}
\epsfig{figure=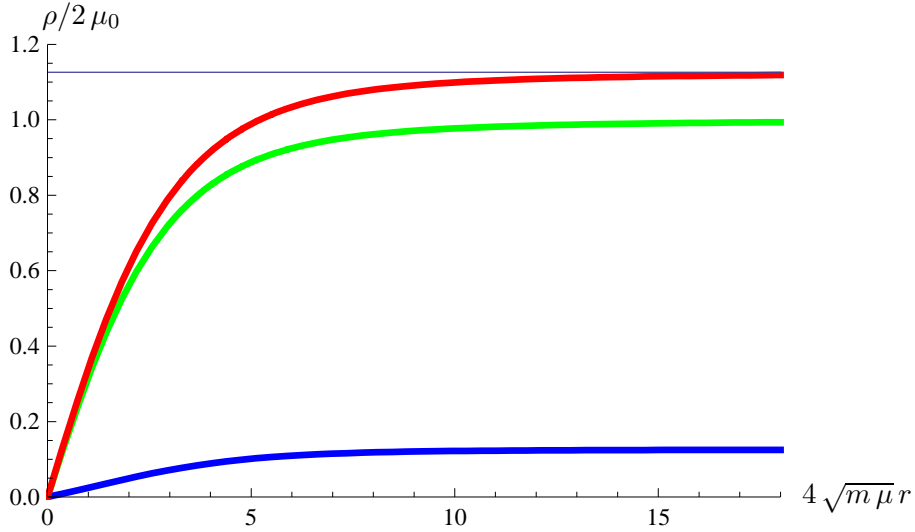, width=.65\textwidth}
\end{psfrags}
}
\caption{The single vortex profile obtained from Lagrangian (\ref{Leff}). The LO piece is in green, the NLO is in blue, their sum
with $\varepsilon=1$ is the red curve. The NLO ($\varepsilon=1$) background value of the order parameter is shown for reference ($\mu_0=\mu/\varepsilon$).}
\label{fig:1vortex}
\end{figure}
In order to relate these results to experimentally relevant quantities, (\ref{Leff}) was used to study $\Phi(x)$ profile in a harmonic trap
where
\ber
\mu\rightarrow\mu\left(1-\frac{r^2}{L^2}\right).
\label{muUx}
\eer
The trap sets the scale, so, even the LO derivative term is suppressed by $L_{\mu}^2/L^2 \sim N^{-2/3},$ where $N$ is number of particles in the trap. Derivative term 
becomes relevant only near the boundary of the cloud where $L\sim (m\,(\mu-{\rm U}(\vec{x})))^{-1/2}$ \cite{Son:2005rv}. This illustrates validity of 
Thomas-Fermi approximation, a well known fact. Then, using Thomas-Fermi solution, $\Phi(r)=\phi_0\left(\mu\left[1-{r^2}/{L^2}\right]\right),$
where $\phi_0(\mu)$ is given in (\ref{phi0}), one expresses the size of a vortex in terms of the trap size and particle
number. In the spherical trap case we get
$r_0 \simeq 0.20(1+\varepsilon\,0.27)\,L\,(N\,\varepsilon)^{-1/d}.$ In the cylindrically symmetric case
$r_0 \simeq 0.23(1+\varepsilon\,0.21)\,(\varepsilon\,{\rm V}/{N})^{1/d}$ where ${\rm V}$ is the total volume of the trap. The predictions for $d=3$ are then
\ber
r_0 &=& 0.25\,L\,N^{-1/3},\nonumber \\
r_0 &=& 0.28\,({\rm V}/{N})^{1/3}
\label{r0d3}
\eer
for the spherical and cylindrical traps, respectively, where ${\rm V}$ is the total volume of the cylindrical trap.
The two configurations are the limiting cases which span shapes of the trapping potentials realized in the experiments.

\section{Normal-Superfluid Interface in the Polarized Unitary Fermi gas}

In general, derivative terms are relevant for any inhomogeneous phenomenon with typical length scale of order of inter-particle distance. 
Such a situation arises in description of the structure of the superfluid-normal phase interface encountered in the trapped
imbalanced gas ($\delta\mu\neq0$) \cite{zwierlein-2006-311,partridge-2005,partridge-2006-97}.

\begin{figure}[t]
\centering{
\begin{psfrags}
\psfrag{b}{${\rm V}\,(2\,\pi)^{d/2}/\mu_0(\mu_0\,m)^{d/2}$}
\psfrag{a}{$\phi/\mu_0$}
\epsfig{figure=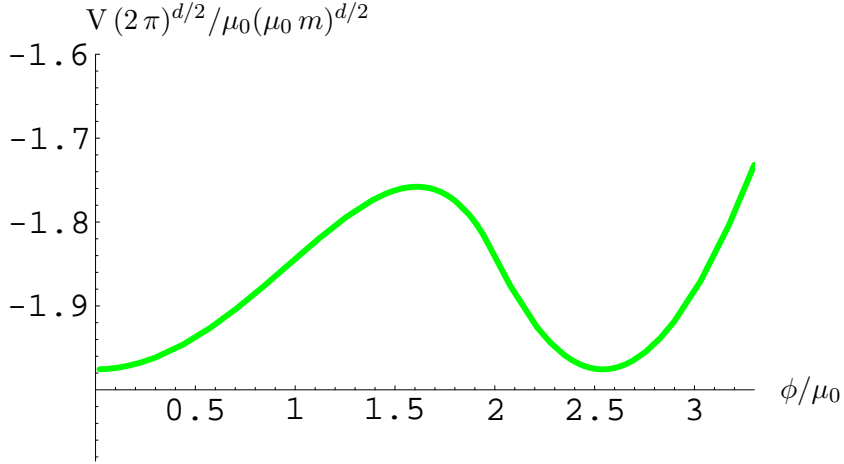, width=.65\textwidth}
\end{psfrags}
}
\caption{The NLO effective potential of the bulk medium ${\rm V}$ at $\delta\mu\simeq\delta\mu_c$ as a function of the order
parameter $\phi$ ($\mu_0=\mu/\varepsilon$).
}
\label{fig:pot_deltamu}
\end{figure}

\begin{figure}[t]
\centering{
\begin{psfrags}
\psfrag{b}{$\Phi(z)/\mu_0$}
\psfrag{a}{$4\,\sqrt{m\,\mu}\,z$}
\epsfig{figure=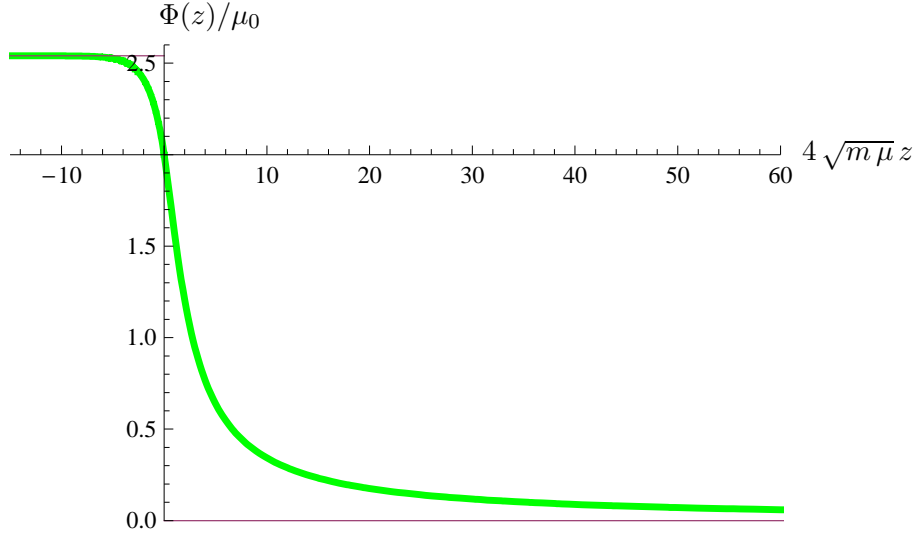, width=.65\textwidth}
\end{psfrags}
}
\caption{The superfluid-normal phase interface from the NLO effective Lagrangian with $\varepsilon=1.$}
\label{fig:surface_tension}
\end{figure}

Shown in Fig.~(\ref{fig:pot_deltamu}) is the NLO effective potential of the bulk medium ${\rm V}$ at the critical polarization 
\cite{Rupak:2006et,Nishida:2006eu}. Using effective Lagrangian for the polarized unitary Fermi gas (\ref{Leffpolreal}) one is 
to solve for the order parameter spatial profile connecting superfluid and normal minima of the potential. Unfortunately, due 
to flatness of the LO potential \cite{Rupak:2006et,Nishida:2006eu} the interface does not exist at the LO in $\varepsilon$ and 
the problem is partially non-perturbative. Let us determine the structure of the superfluid-normal phase domain wall using Lagrangian 
(\ref{Leffpolreal}) with $\varepsilon$ set to 1. Note that $\mathcal{L}_{eff}[\Phi,\mu,\delta\mu]$ is not smooth as indicated by the 
presence of the $\Theta$ function terms. This appears to be an artifact of the $\varepsilon$ epsilon expansion which treats the 
chemical potential as a perturbation. Shown in Fig.~(\ref{fig:surface_tension}) is the resulting profile $\Phi(z).$ It has a small 
discontinuity in its derivatives at $\Phi(z)=2\,\mu_0;$ its width is of the order of the inter-particle separation, so the
derivative expansion used to approximate effective action (\ref{Seff}) with the effective Lagrangian (\ref{Leffpolreal}) is valid in 
this case. Since we used $\mathcal{L}_{eff}[\Phi,\mu,\delta\mu]$ at $\varepsilon=1,$ the location of the superfluid minimum and the critical value of 
$\delta\mu$ are slightly different from the results found in the perturbation theory \cite{Rupak:2006et,Nishida:2006eu}.
The surface free energy per unit area is then
\ber
\sigma&=&-\int{\rm d}z\,
\left(\mathcal{L}_{eff}[\hat{\Phi}(z),\mu,\delta\mu_c]-\mathcal{L}_{eff}[\hat{\Phi}(z=\infty),\mu,\delta\mu_c]\right)=
0.58\,\sqrt{\frac{\mu_0}{m}}\,\left(\frac{\mu_0\,m}{2\pi}\right)^{d/2}+\nonumber \\
&&+{\cal{O}}(\varepsilon^2),\qquad \mu_0=\frac{\mu}{\varepsilon},
\label{surface_energy}
\eer
where $\hat{\Phi}(z)$ is the solution shown in Fig.~(\ref{fig:surface_tension}).
In the case of polarized gas in a spherical trap
\ber
\int_{S}\sigma \simeq \mu_0(\mu\,m)^{d/2}\,L^{d}\int_{\tilde S}\frac{1}{2}{\tilde n}^{\frac{d+1}{d}}
0.33\frac{1}{\sqrt{m\mu}L},
\label{surface_energy_trap}
\eer
where $\tilde n=n/(\mu_0\,m)^{d/2}$ is the (dimensionless) density on the superfluid side of the interface; $\int_{\tilde S}$ is the (dimensionless) integral over the interface. Note that the total surface energy is in the units of $\mu(\mu\,m)^{d/2}\,L^{d}$ which is the units
of the bulk free energy. 

In the Thomas-Fermi approximation location of the superfluid-normal phase boundary in a harmonic trap is determined by
\ber
\delta\mu=\delta\mu_c(\vec{x}) \equiv 2\,\mu_0(\vec{x})\,(1-\beta\,\varepsilon)+{\cal{O}}(\varepsilon^2),
~\beta \simeq 0.47,\,\mu_0(\vec{x})=\frac{\mu}{\varepsilon}\left(1-{r^2}/{L^2}\right)
\label{deltamuc}
\eer
\cite{Rupak:2006et,Nishida:2006eu} which means that 
the interface is located at the distance
\ber
{R}={L}\sqrt{1-\frac{\delta\mu}{2\,\mu_0}}\left(1-\varepsilon\frac{\beta\,\delta\mu}{4\mu_0-2\,\delta\mu}\right),~~\mu_0=\frac{\mu}{\varepsilon},
\label{Rinterface}
\eer
from the center.
To calculate $1/{\sqrt{m\mu}\,L}$ we use Thomas-Fermi solution for the order parameter of polarized gas in a spherical trap
\ber
\Phi(r)=\phi_0\left[\mu\left(1-\frac{r^2}{L^2}\right)\right]\Theta(R-r),
\label{phir_polar_TF}
\eer
where $\phi_0(\mu)$ is the homogeneous order parameter (\ref{phi0}) and the superfluid core radius, $R,$ has been defined in (\ref{Rinterface}).
The result is almost independent of polarization,
$\delta=({\rm N_{\uparrow}-N_{\downarrow}})/({\rm \rm N_{\uparrow}+N_{\downarrow}});$ for $\delta=0.53$ we
get
\ber
\frac{1}{\sqrt{m\mu}\,L} &\simeq& \left(\varepsilon\,N\right)^{-1/d}\,0.47(1+0.21~\varepsilon)
\label{muLspher}
\eer
which for $d=3$ gives
\ber
\int_{S}\sigma &\simeq& \mu(\mu\,m)^{3/2}\,L^{3}\,\int_{\tilde S}\frac12{\tilde n}^{{4}/{3}}\,0.19~N^{-1/3}.
\label{sigma_spher}
\eer
For $N=10^5$ the coefficient is 0.004 which is four times the value of \cite{desilva-2006-97}.

\section{Summary}
We have calculated effective Lagrangian of the unitary Fermi gas
using $\varepsilon$ expansion technique as well as derivative expansion. The leading order 
derivative terms have been determined by gauging the $U(1)$ particle number symmetry and requiring gauge invariance of the theory.
We have argued that in some realistic situations the derivative and $\varepsilon$ expansions are related: for large enough scale phenomena so that the typical length scale is not 
smaller than $L_{\mu} \sim (m \mu)^{-1/2}$ 
the LO derivative terms are ${\cal O}(1),$ while the higher 
order derivative terms are at least ${\cal O}(\varepsilon^2)$ 
and, so, are to be neglected when working to NLO.
Quasi-locality of the effective action functional (${\it i.}{\it e.}$ validity of the derivative expansion)
is due to the feature of the $\varepsilon$ expansion that $\mu \sim \varepsilon$ and is, therefore,
treated as a perturbation.
In this approach to inhomogeneity description one has to deal with differential equations (equations of motion for the order parameter)
which is simpler than the usual Bogoliubov-de Gennes method \cite{Degennes:1966fr} requiring
having to deal with non-local effective action functional.

As an application of the effective Lagrangian obtained, we have determined the structure of a vortex in the unitary Fermi gas, 
and the superfluid-normal phase interface profile, all to the NLO in $\varepsilon.$ 
Predictions for the vortex size in spherical and cylindrical traps are shown in (\ref{r0d3}); surface free energy results are given by Eqs. (\ref{surface_energy},\ref{sigma_spher}).

Some other applications of the functional include
\begin{itemize}
\item
{study of multiple vortex configurations and calculation of the parameters of the vortex
lattices observed in the unitary Fermi gas;} 
\item 
{study of properties of strongly interacting Fermi gas in a periodic potential (optical lattice), a subject of recent experiments \cite{Zwerger:2007ab};}
\item{recently new Monte Carlo simulations of trapped few fermion system at unitarity have been performed
\cite{Bertsch:2007br,blume-2007,vonstecher-2007}. Assuming validity of continuous medium description, one should consider the case $L_{trap} \sim (m\,\mu)^{-1/2}.$ Thomas Fermi
approximation is no longer valid, while derivative expansion based functional (\ref{Leff}) should still be valid. 
For systems with fewer particles an alternative technique such as in, for example, \cite{Nishida:2007pj} must be employed.}
\end{itemize}
\section{Acknowledgment}
I would like to thank Chuck Horowitz, Yusuke Nishida and Dam Son for useful conversations and Aurel Bulgac for useful conversations and commenting on the manuscript. This
work is supported in part by the US Department of Energy grants DE-FG02-87ER40365 and DE-FG02-05ER41375 (OJI) and by the National Science Foundation grant PHY-0555232.
\bibliography{cfl}
\end{document}